\documentclass{emulateapj}
\usepackage{graphicx}

\def\see{\mbox{$^{\prime\prime}$}}
\shorttitle {The earliest galaxies and the epoch of reionization }
\shortauthors{Pentericci et al.}

\begin{document}
\title{New observations of z$\sim 7$ galaxies: evidence for a patchy reionization}
\author{L. Pentericci\altaffilmark{1},
E. Vanzella\altaffilmark{2}, 
A. Fontana\altaffilmark{1},
 M. Castellano\altaffilmark{1},
T. Treu\altaffilmark{3},
A. Mesinger\altaffilmark{4}, 
M. Dijkstra\altaffilmark{5}, 
A. Grazian\altaffilmark{1}, 
M. Brada\v{c}\altaffilmark{6},
C. Conselice\altaffilmark{7},
S. Cristiani\altaffilmark{8}, 
J. Dunlop\altaffilmark{9}
A. Galametz\altaffilmark{1},
M.  Giavalisco\altaffilmark{10}, 
E. Giallongo\altaffilmark{1},
A. Koekemoer\altaffilmark{11},
R. McLure\altaffilmark{9},
R. Maiolino\altaffilmark{12}, 
D. Paris\altaffilmark{1}, 
P. Santini\altaffilmark{1}
}

\affil{$^1$ INAF Osservatorio Astronomico di Roma,  Via Frascati 33,00040 
Monteporzio (RM), Italy}
\affil{$^2$ INAF Osservatorio Astronomico di Bologna, Italy}
\affil{$^3$ Department of Physics, University of California, Santa Barbara, CA 93106-9530, USA}
\affil{$^4$ Scuola Normale Superiore,Piazza dei Cavalieri 7, I-56126 Pisa, Italy}
\affil{$^5$ Max-Planck-Institut fur Astrophysik, Karl-Schwarzschild-Str. 1, 85741 Garching, Germany}
\affil{$^6$ Department of Physics, University of California, Davis, CA 95616, USA}
\affil{$^7$ School of Physics \& Astronomy, University of Nottingham, Nottingham NG7 2RD, UK }
\affil{$^8$ INAF Osservatorio Astronomico di Trieste, Via G.B.Tiepolo 11, 
34131 Trieste, Italy}
\affil{$^9$ Institute for Astronomy, University of Edinburgh, Royal Observatory, Edinburgh EH9 3HJ, UK }
\affil{$^{10}$Department of Astronomy, University of Massachusetts, Amherst, MA 01003, USA }
\affil{$^{11}$ Space Telescope Science Institute, 3700 San Martin Dr., Baltimore, MD 21218, USA   }
\affil{$^{12}$ Kavli Institute for Cosmology, University of Cambridge, Madingley Road, Cambridge, CB3 0HA, UK}

\begin{abstract}

We present new results from our search for $z\sim 7 $ 
galaxies from deep spectroscopic observations of candidate z-dropouts
in the CANDELS fields.  Despite the extremely low flux limits achieved 
 by our sensitive observations, only 2 galaxies have robust redshift identifications, one from its Ly$\alpha$ 
emission line  at z=6.65, the other from its Lyman-break, i.e. the continuum discontinuity at the Ly$\alpha$ wavelength  consistent with a redshift 6.42, but with no emission line. In addition, for $23$ galaxies we
present deep limits in the Ly$\alpha$ EW derived from the non detections in ultra-deep observations.

Using this new data as well as previous samples,  we assemble a total  of 68
candidate z$\sim 7 $ galaxies with deep spectroscopic  observations, 
of which 12 have a line detection. With this much 
enlarged  sample we can  place solid   constraints on the declining fraction of Ly$\alpha$ 
emission in z$\sim$7 Lyman break  galaxies compared to z$\sim 6$, both for bright and faint galaxies. 
Applying a simple analytical model,  we show that the present 
data favor a patchy  reionization process rather than a smooth one. 
\end{abstract}

\section{Introduction}
The use of Ly$\alpha$ transmission by 
the intergalactic medium (IGM) as a probe of its ionization
state during the reionization epoch has been proposed already many years ago (Miralda-Escude \& Rees 1998; Santos et al. 2004).
Strong Ly$\alpha$ emission, powered by star formation  
is present is many distant galaxies: being a resonant line, it is sensitive to even small 
quantities of neutral hydrogen in the IGM, and it is easily suppressed \citep{loeb,malho,zheng}.
We thus expect the observed properties of Ly$\alpha$ emitting galaxies to change 
at higher redshifts,  when the IGM becomes more neutral.
A very common approach for studying the reionization history of the Universe using Ly$\alpha$ emitting galaxies is
to determine the evolution of the Ly$\alpha$ luminosity function and the clustering properties 
of narrow band selected Ly$\alpha$ emitters (LAEs e.g. Ota et al. 2008, Ouchi et a. 2010; Kashikawa et al. 2011, 
Clements et al. 2012, Faisst et al. 2014). A recent complementary  approach and the one used in this paper,  is 
instead  to  measure the redshift evolution of the Ly$\alpha$ fraction in Lyman break galaxies (LBGs), 
i.e. the percentage of LBGs that have an appreciable Ly$\alpha$ emission line (e.g. Stark et al. 2010). 
Indeed this fraction is supposed  to increase, 
as we move to higher redshift since galaxies are increasingly  young 
(hence with stronger intrinsic Ly$\alpha$) and  almost dust free (Finkelstein et al. 2012)
 which facilitates the escape of Ly$\alpha$ photons.
On the other hand this fraction is expected to fall-off as we approach 
the time when  the intergalactic medium becomes significantly neutral  
and the galaxies' Ly$\alpha$  emission is progressively attenuated. 
Compared to other probes of reionization such as the evolution of the  LAE luminosity function 
this approach can overcome concerns about  intrinsic density evolution  of the underlying population (Stark et al. 2010). 

Intriguingly, early measurements with this technique
suggest a strong drop in the Ly$\alpha$ fraction near z$\sim$7,  more significant for relatively fainter galaxies.
In particular in a series of recent works, a lack
of Ly$\alpha$ emission was found at $z\sim 7$  compared to $z\sim 6$ by several independent teams: in our previous observations (Pentericci et al. 2011; Vanzella et al. 2011; Fontana et al. 2010; P11, V11 and F10 from here on)
we found 4  Ly$\alpha$  emitting galaxies (plus another object with a tentative line detection) 
out of a sample of 20 robust candidates.
Similar or lower fractions were found by \cite{she12},\cite{car12}, 
Brada\v{c} et al. (2012), Ono et al. (2012)  although considerable field-to-field variations are present due to the small 
number of candidates observed in each sample (see for example Figure 8 in Ono et al. 2012).
In our favored interpretation,  the lack of line emission is  due to a substantial increase in the
neutral hydrogen content  of the Universe
in the time between $z\sim 6$ and $z\sim 7$. 
Comparing our data to the predictions of the semi-analytical models by  Dijkstra et al. (2011)
we concluded that to explain the observations a substantial change of the neutral hydrogen fraction of the order of $\Delta  \chi_{H I} \sim 0.6$ in a time $\Delta z \sim 1$ was required, assuming that the galaxies' physical 
properties  remain constant during this time.  Recent observations pushing to $z\sim8$ are consistent with this interpretation (Treu et al.\ 2013; Schmidt et al. 2014).
\\
However other factors could also play a  role in the Ly$\alpha$ quenching: in particular we cannot rule out the possibility that  a change in some of the intrinsic galaxy properties (the Lyman continuum escape fraction, wind properties  and  dust content) could at least partially contribute to the lack of Ly$\alpha$ emission.
Indeed the  interpretation of the results as only due to the change in the neutral hydrogen fraction 
was questioned by several successive works (e.g. Jensen et al. 2013, Forero-Romeiro et al. 2012, Bolton \& Haehnelt 2013, Taylor \& Lidz 2014, Dijkstra et al. 2014). 
 In particular Bolton \& Haehnelt (2013) suggested  that the opacity of the intervening IGM red-ward of rest-frame Ly$\alpha$ can rise rapidly in average regions of the Universe simply because of the increasing incidence of absorption systems, which are optically thick to Lyman continuum photons.
They claimed that the data do not require a large change in the IGM neutral 
fraction  from z$\sim$6 to $\sim$7. However such a rapid evolution of the photo-ionizing background could be very difficult to achieve without requiring either a late reionization, or an
emissivity at $z < 6$ which is too high to be consistent with observations of the Ly$\alpha$
forest (e.g. Sobacchi \& Mesinger 2014).
Preliminary estimates suggest that the neutral fraction constraint 
relaxes only mildly when taking into account the absorption systems (Mesinger et al. in preparation).
\\
We also mention the very recent work by  Taylor \& Lidz (2014), pointing  out that  sample variance is not negligible for existing surveys: considering the large spatial fluctuations of the medium owing to an inhomogeneous reionization,  the required neutral fraction at z$\sim$7 can somehow be reduced to less extreme values. Indeed 
the observational results are presently based on small data-sets, 
with considerable field to field variations (Pentericci et al. 2011), and mostly focusing on 
the brightest candidates ($M_{UV} < -20.5$). 

The complex  topology of reionization is also a  highly  debated matter.
Depending on the nature of  the main sources of reionization,
it is expected that the characteristic scale of the reionization process might change substantially  (Iliev et al. 2006; Furlanetto et al. 2006). Accurate theoretical predictions for
the morphology and sizes  of H II regions depend on the abundance and clustering of the ionizing sources themselves, in addition to the underlying inhomogeneous density field and clumpiness  of the gas in the IGM (McQuinn et al. 2007, Sobacchi\& Mesinger 2014).
Observations of Ly$\alpha$ emitting galaxies and their  clustering have the potential to reveal the signature of patchy reionization, although early results have been inconclusive (Kashikawa et al. 2011, Ouchi et al. 2010)
 
In this  paper we present new observations of z$\sim 7$ candidates, significantly increasing the statistics of previous works (especially in the faint regime, thanks to the inclusion of lensed candidates), which will allow us to  assess the emergence of Ly$\alpha$ emission  at high redshift with greater accuracy and address some of the above issues. In Section 2 we present the new  observations and the previous data available;
in Section 3 we describe the  simulations used to accurately evaluate the 
 sensitivity  of our spectroscopic observations. In Section 4 we first 
evaluate the new limits on the Ly$\alpha$ fractions at high redshift, and we derive the neutral hydrogen fraction that is needed to explain the observed decrease; then applying a simple phenomenological model we derive new  constraints on the topology of reionization. In Section 5 we summarize our findings.

All magnitudes are in the AB system, and we adopt
$H_0=70$~km/s/Mpc, $\Omega_M=0.3$ and $\Omega_{\Lambda}=0.7$.

\section{Observations }
In this section we summarize the new observations presented in this work as well as previous data that we will use in this paper.
\subsection{UDS field}
We  selected candidate z$\sim$7 galaxies in the UDS field
from CANDELS multi-wavelength observations (Galametz et al. 2013). 
 Objects were detected in the J band  and then the color selection criteria presented by Grazian et al. (2012).
were applied\footnote{Note that these observations were performed before the official CANDELS catalog was released.
Therefore at the time of mask preparation we adopted the J-detected catalog already used 
in Grazian et al.  and in other works.}
Observations were taken in service mode with the FORS2 spectrograph on
the ESO Very Large Telescope. We used the 600Z holographic grating, that provides the highest
sensitivity in the range $8000-10000$\AA\  with a spectral resolution
 $R\simeq 1390$ and a sampling of 1.6\AA\  per pixel for a 1\see~slit.
Out of the entire sample of z-dropout candidates (which consists of 50 galaxies), we 
placed  a   total of 12 galaxies in the slits (the selection was just driven by the geometry of the mask). The rest of the mask was filled with i-dropouts (Vanzella et al. 
in preparation) and other targets  such as massive high redshift galaxies and AGN candidates.
The sources have been observed through slitlets 1\see~wide by 10-12\see~long.
The observation strategy was identical to the one adopted in P11 and previous papers: 
series of spectra were taken at two different positions, offset
by $4\farcs$ (16 pixels) in the direction perpendicular to the dispersion.
The total net integration time was 
15.5 hours for each object.
Data were reduced using our dedicated pipeline which was described in detail in F10 and V11.
Here we only mention that our pipeline performs the sky-subtraction as typically done for the 
near-IR, subtracting  the sky background between two consecutive exposures, exploiting the fact that the target spectrum 
is offset due to dithering in the classic ABBA pattern. 
Our algorithm implements a "A-B" sky subtraction joined with a zero
(e.g., median) or first order fit of the sky along columns that
regularized possible local differences in the sky counts among
the partial frames before they are combined. We find that this procedure ensures the best final results
when searching for faint emission lines, especially in the reddest part of the spectra  
where many strong skylines are present. 
The two dimensional sky-subtracted partial frames are also combined (in the pixel domain) to produce
the weighted RMS map, associated to the final reduced spectrum. 
This allows us to calculate the two dimensional signal to noise spectra, useful to access the reliability of the spectral features
Finally we also take extra care in the alignment of the different frames before the 
combination.
\\
For one of the candidates with a relatively bright continuum magnitude (J=25.98),  UDS29249, 
we detect a  faint continuum emission  in the red part of the spectrum 
beyond  $\sim $9100 \AA\ as shown in Figure 1.  The total integrated S/N of the flux is $\sim 10$; while the detection is 
relatively secure in the wavelength range 9160-9240 \AA,  the exact position of the break 
(that is ascribed to the IGM) is difficult to locate 
both because of the faintness of the emission and also because of the residuals of the bright  sky emission lines
 in the region immediately below 9100\AA.
We conservatively estimate  a redshift that ranges between 6.31 and 6.53 (in the table we report $6.42\pm0.11$).
Note that this is perfectly consistent with the photometric redshift distribution obtained from the CANDELS photometry, also shown in Figure 1. 
Recently Wilkins et al.  (2014) discussed the possibility that a newly identified Y-dwarf population, as well as the late
 T-dwarfs stars  might contaminate the photometric selection  and spectroscopic follow-up of faint and distant galaxies (see also Bowler et al. 2014).  
Our target appears very compact but still resolved in the HST J-band (as the majority of the $z\sim 7$ candidates); in addition  its colors are not consistent with  those of Y and T dwarf. If we place the target in the  
$z-Y$ vs $Y-H$  plot, as in Figure 3 of Wilkins et al., 
the object is almost coincident with the high-z star-forming galaxy track, and  very distant from the position 
of both the L- and T-dwarf spectral standards as well as  the tracks of the model Y-dwarfs.
Thus this gives us extra confidence that this is a true high redshift galaxy 
without Ly$\alpha$ emission in its spectrum.
\\
All other candidates are undetected (meaning that no feature is detected).
In Table 1 we report the candidates, RA and Dec, their J band magnitudes  and the limiting EW. 
For the undetected objects we assume a redshift of 6.9, which is the median redshift
of the selection function (see Grazian et al. 2012).

\subsection{ESO Archive}
We searched the ESO archive  for observations of high redshift objects;  
in particular we retrieved  the data from the observations carried out within the 
 program ESO 088.A-1013 (PI Bunker). This program used  
the same observational setup used  above, with a total net integration time of 27 hours.
It observed a mixture of z and i-dropouts. 
Recently the results were presented by \cite{caru2013} and 
the authors report all non detections for the candidate z$\sim 7$ galaxies.
\\
Caruana et al. have observed candidate high redshift galaxies selected in 
previous works
(Bouwens et al. 2011, McLure et al. 2010  and Wilkins et al. 2011) which used each a different selection criteria, 
from different color-color cuts to photometric redshifts.
Since we want to work on a sample that is as homogeneously selected as possible, we have selected our own list of z-dropouts
again using the color criteria presented by  Grazian et al. (2012). We then cross correlated our list
with the targets in \cite{caru2013}.
We found 9 matching objects, of which 5 are in common with the sample already observed in F10.  We then  retrieved the raw (public) data from the ESO archive and then processed through our own  
pipeline (V11)  as all other data in this work.
Here we present the results for the 4 new targets that are not in common with F10 (see Table 1). Further 
results in particular   the extremely deep  combined  spectra of the objects in common between the Caruana et al. program and  F10 (52 hours) will be presented elsewhere (Vanzella et al. in preparation).
 \\
The results are again presented in Table 1. We detect a significant emission line in one of the 4 new objects, galaxy \# 34271 
in the  GOODS-South field corresponding to galaxy ERSz-2225141173 in \cite{caru2013}. 
The line is detected at 9301 \AA\ and shows the typical asymmetry of Ly$\alpha$ which would place the object at redshift $6.649\pm0.001$. In Figure 2 we show the 1-dimensional and 2-dimensional spectra of the galaxy.
The EW of the line is 43 \AA, calculated from its measured line flux and the Y band magnitude from the GOODS CANDELS catalog. 
\\
As we mentioned above, \cite{caru2013} report no emission line for this galaxy, but a median EW limit of 28.5\AA.
We ascribe the difference to  the fact that for most of the data reduction steps the authors used the standard ESO pipeline, 
while we use our own  pipeline that has been  tailored specifically  to the detection of faint high redshift emission lines (see the description above).
The other 3  targets are undetected and we can set stringent limits of their
Ly$\alpha$  EW limit ranging from $\sim 10$\AA\ to $\sim 25$\AA\ depending on the continuum magnitude
 and assuming that they are placed  at redshift 6.9. 
Obviously the actual EW limit depends sensibly on the exact redshift  of the objects (see F10 Figure 1).

\begin{table*}
\scriptsize
\begin{center}
\caption{Spectroscopic properties of observed z-dropouts in the new  fields \label{tbl-2}}
\begin{tabular}{cccccccccc}
UDS \\
\hline
ID    & RA   & Dec & J125 &  & $M_{UV}$ &  z  &  EW(S/N=5)   \\ 
\hline
29249 & 34.226135  &	–5.1510921  & 25.985 & & -20.93  &   $\sim$6.42$\pm 0.11$ & $<$9                   \\
28737 & 34.229103  &	–5.1533098  & 25.967 & & -20.95  &  -- &$<$12                  \\
16910 & 34.226192  &	–5.2033339  & 26.503 & & -20.42  & -- &  $<$20      \\      
23427 & 34.298386  &	–5.1760311  & 25.826 & & -21.09  & -- & $<$10    \\               
15399 & 34.233883  &	–5.2100158  & 25.426 & & -21.49  & -- & $ <$7     \\               
16119 & 34.253719  &	–5.2068028  & 26.297 & & -20.62  & -- & $<$16        \\           
16669 & 34.279049  &	–5.2043710  & 26.479 & & -20.44  & -- & $<$19        \\           
16974 & 34.313725  &	–5.2030821  & 26.02  & & -20.87  & -- & $<$12        \\             
16094 & 34.3180048 &    -5.2069350  & 27.006 & & -19.91  & -- & $<$30            \\
14435 & 34.323608  &	–5.2141371  & 26.570 & & -20.35  & -- & $<$20      \\             
8912  & 34.2815881 &	-5.23757910 & 26.805 & & -20.11  & -- & $<$25        \\     
12402 & 34.3203425 &	-5.22268940 & 27.013 & & -19.90  & -- & $<$30              \\  
\hline
GOODS-S\\
\hline
ID     & RA & Dec  &  J125 & &  $M_{UV}$ &   z   &  EW  \\
\hline   
20439 & 53.09556 & -27.73609  & 27.11 &  & -19.81 & --  &   $<$25     \\ 
24805 & 53.11627 & -27.6845   & 26.08 &  & -20.84 & --  &   $<$10    \\ 
14259 & 53.16164 & -27.78533  & 27.08  & & -19.84 & --  &   $<$25  \\
34271 & 53.09377 &- 27.68814  & 27.44 &  & -19.48 & 6.65&   43   \\
\\
\hline
Bullet \\
\hline
ID &  RA & Dec & J110  &    J110 intrinsic  &  $M_{UV}$ intrinsic & z  &  EW  \\
\hline
1   &  104.65470&	–55.974464 & 26.88  &    28.45 & -18.52 & --    &  $<$25        \\
2   &  104.65527&	–55.971901 & 27.03  &    29.05 & -17.92 & --    & $<$29        \\
3   &  	104.66736&	–55.968067 & 25.43  &    28.13 & -18.84 & --    &   $<$7       \\
4   &   104.66375&	–55.928802 & 26.88  &    27.97 & -19.00 &  --   & $<$25       \\
5   &   104.63437&	–55.978603 & 25.98  &    26.78 & -20.19 & --    &  $<$11       \\
6   &   104.62446&	–55.951065 & 25.87  &    28.38 & -18.59 & --    & $<$10        \\
7   &   104.64304&	–55.964756 & 25.95  &    27.75 & -19.22 & --    & $<$11       \\
8   &   104.64549&	–55.924828 & 26.29  &    27.52 & -19.45 & --    & $<$15        \\
9   &   104.63254&	–55.963764 & 26.46  &    28.05 & -18.92 & --    & $<$17        \\
10  &   104.63015&	–55.970482 & 26.47  &    27.67 & -19.30 & 6.740 &  30          \\

 \end{tabular}
\end{center}
\end{table*}

\subsection{Data from previous literature}
Besides the new data, we  include all previously published spectroscopic data on z-dropouts, in order 
 to assemble the largest possible sample of candidate z$\sim 7 $ galaxies with deep spectroscopic observations.
\\
In particular  we consider: (1) the 20 z-dropouts selected in the GOODS-South, NTTDF and BDF4 fields (Castellano et al. 2010a, 2010b) 
whose observations were carried out  by our groups in P11, and previously presented by V11 and F10. 
Of these,  4 show a convincing Ly$\alpha$ emission line, while the tentative detection of a fifth 
candidate originally shown  in F10
was not confirmed by the combination of our own data with the  deeper observations of \cite{caru2013}
(Vanzella et al. in preparation). (2) The 11 bright z-dropout  observed by Ono et al. 
(2012) in the SDF field  of which 3 have bright Ly$\alpha$ in emission. These candidates were detected 
 in deep Y band observations and selected using color criteria that are very similar to ours.  (3)  A subset of the objects presented by  Schenker et al. (2012).  In particular  we select those galaxies  whose colors are consistent  with the z-dropout selection criteria  used in this paper (note that Schenker et al. also observed Y-band dropouts whose photometric redshifts are $>>$ 7). In total we consider 11 objects of which 2 are detected with Ly$\alpha$.
\\
Overall considering new and previous data, we assemble a sample of 68 z-dropouts that have been spectroscopically 
observed with either VLT, Keck or Subaru down to very faint flux limits.
Note that 46 out of 68  have been observed with exactly the same set-up (with FORS2@VLT using grism 600z).
 
\begin{figure}
\epsscale{1.0}
\includegraphics[width = 9cm,clip=]{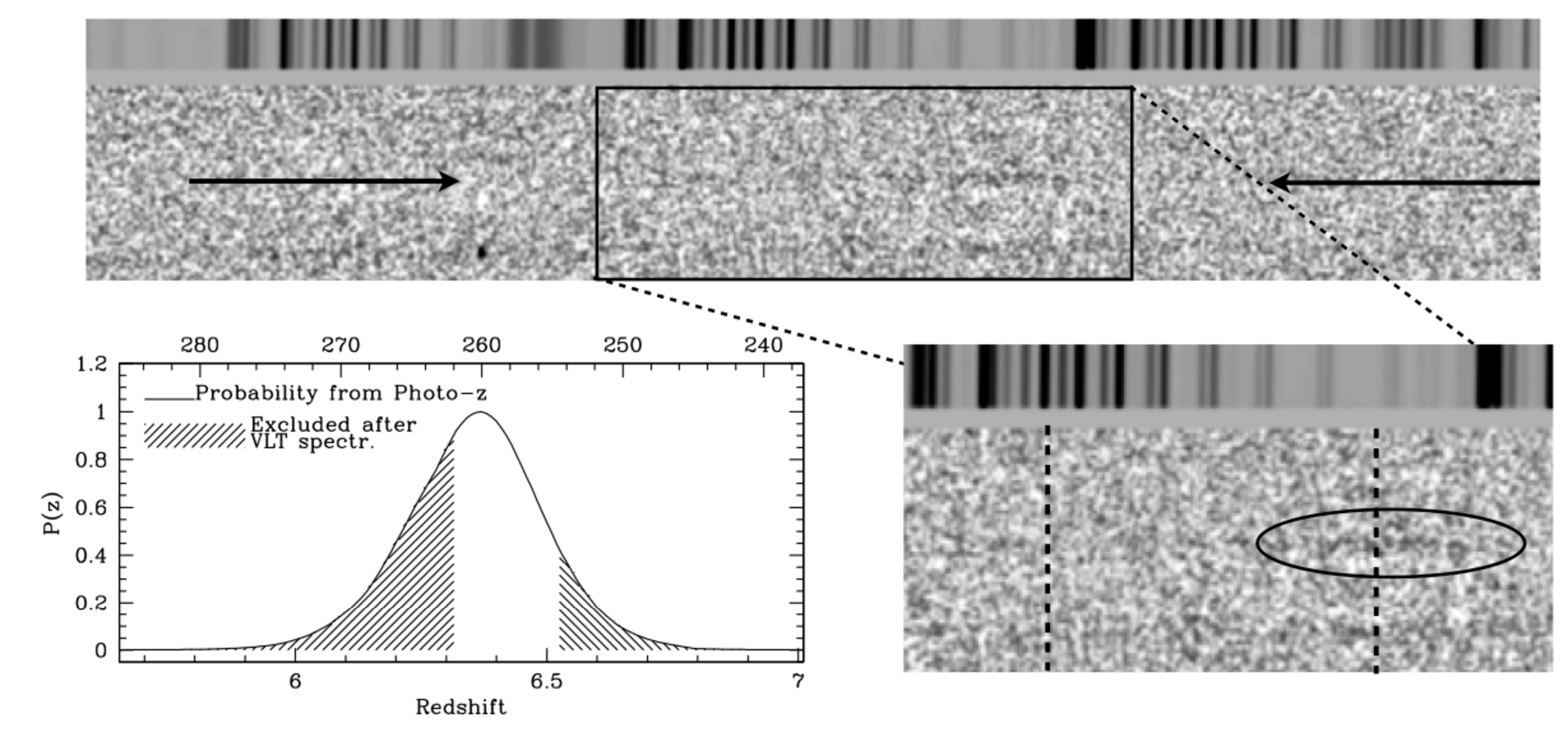}
\caption{The upper panel shows the 2-dimensional spectrum of galaxy UDS29249. The lower right panel in an enlarged view of the wavelength range where we detect faint continuum, with an integrated S/N $\sim 10$. The lower left panel shows the probability distribution function of the photometric redshift, obtained from the CANDELS photometry, which is further restricted when combined to  the spectroscopic observations. In the upper panel note the presence of a serendipitous emission line to the lower left, coming form a redshift 6.058 galaxy.  }
\label{udsfig}
\end{figure}

\begin{figure}
\epsscale{1.0}
\includegraphics[width = 9cm]{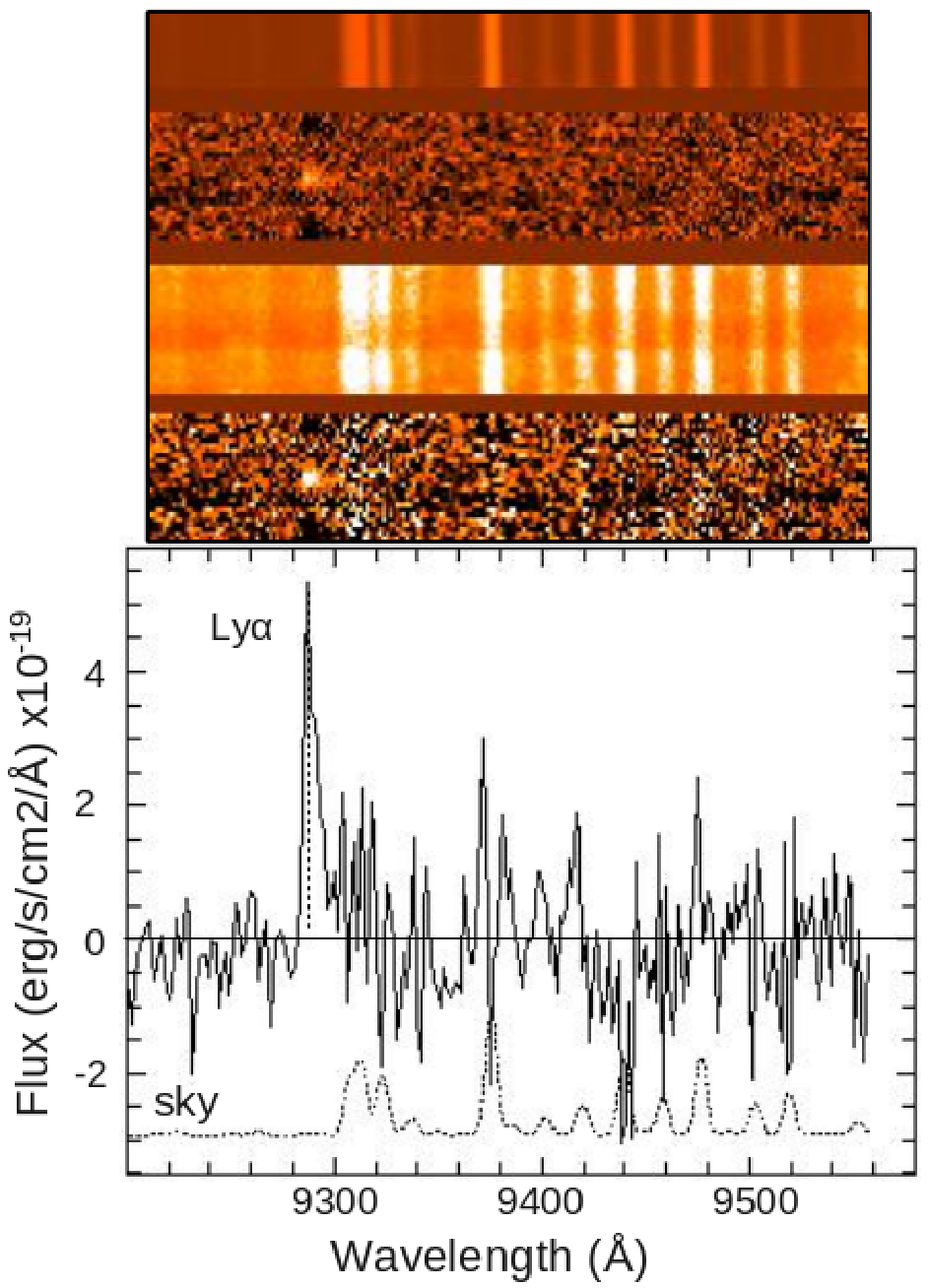}
\caption{The upper panels show the 2 dimensional spectrum of the galaxy GDS34271 (ERSz-2225141173 in Caruana et al. 2013) showing the Ly$\alpha$ line at 9300\AA.
From top to bottom the panels represent the sky emission, the s/n spectrum, the total RMS and the reduced spectrum. The bottom panel is the 1-dimensional extracted spectrum after smoothing by two pixels. }
\label{starkfig}
\end{figure}

\subsection{Bullet cluster}
Brada\v{c} et al. (2012) observed the lensed z-dropouts detected behind the Bullet cluster and selected 
by Hall et al. 2011. The observations were carried out with FORS2@VLT using 
the same observational set-up as  in our various programs
(V11, P11), and above (UDS and GOODS-S)  with a total net  integration time of 16.5 hours.
 Data were reduced using our own pipeline (V11). 
\\
The confirmation of one galaxy showing Ly$\alpha$ emission consistent with a redshift of
 6.74 was presented by Brada\v{c} et al. (2012). 
Here we also consider the observations and limits, in terms of Ly$\alpha$ line detection  of the rest of the sample.
In Table 1 we report the resulting limits  on the Ly$\alpha$ EW for each galaxy,
assuming  again a median redshift of 6.9. 
However note that in this case the median expected redshift of the sample is  $> 7$:
the  redshift probability distribution function of these candidates is much larger and 
extends well beyond z=8, differing considerably 
from the other samples presented here  (for reference see Figure 5 in Hall et al. 2012). 
This is due to the fact that the candidates were selected by applying criteria based on a $z_{850}-J_{110}$ color 
(due to the nature of the HST data available).
In the following section  we will consider, whenever necessary,  the appropriate selection function of this sample.

\section{Simulations}
To determine the EW limit achieved by our observations for each of our targets, and reported in Table 1,  we performed detailed  2D simulations that we briefly describe here.
We take real individual  MXU raw  frames, corresponding to slits where no targets were  detected  and insert
an emission line of a given flux, at a given wavelength and  at a given spatial position corresponding to the middle of the slit.  
The emission line is modeled as a Gaussian that is then truncated 
to half to simulate the typically asymmetric emission lines that are 
routinely observed at lower redshift, and then further convolved  with the seeing 
(with values varying from 0.6\farcs to 1.2\farcs as in the real observations).
The individual frames are then processed
as normally done during the reduction procedure: after
standard flat-fielding, we remove the sky emission lines by
subtracting the sky background between two consecutive
exposures, exploiting the fact that the target spectrum
is offset due to dithering (ABBA technique). Then spectra 
have been wavelength calibrated (using lamp exposures) and finally, they  are flux-calibrated using the observations of spectrophotometric standards and combined.
\\ 
The resulting 2D frame is then scanned with a window of  $7 \times 5 $ pixel to see if there is a detection, 
and in this case the signal to noise ratio of the line is registered.
The simulations are repeated after  shifting the Ly$\alpha$ emission line along the dispersion axis in steps of $1.6$\AA, and in the end we cover the  redshift range from z=5.68 to z=7.30. At each redshift step, the scan is repeated and the S/N of the line is registered again.
\\
The emission lines are then varied in terms of total flux (from 0.24 to $1.6\times 10^{-17} erg s^{-1} cm^{-2}$), 
and full width half maximum  (varying from 230 to 520 $km s^{-1}$). 
These values are in the range of the real observed Ly$\alpha$ lines.
The entire procedure is repeated for each combination of line flux and FWHM.
The simulations were performed for more than one slit in order to cover the entire CCD (top and bottom chips) of the FORS2 MXU frame 
and the results are always the same to within 5\%.
Although our candidates were placed at the center of the slits in most cases,
we also checked for possible differences by placing the initial emission line at 
various  positions along 
individual slits,  i.e. we shifted the Ly$\alpha$ up and down by a few pixels.
Again no significant difference is found. 
In Figure 3 we show one of the results of these tests, with the three colored curves 
(red, green and black)
representing the resulting S/N of a line with flux  1.6 $\sim 10^{-17} erg s^{-1} cm^{-2}$, positioned in three different 
slits: for each slit the result is  the average of 3 different positions  along the spatial axis (0,+5 and -5 pixel). It is evident that the differences in the resulting S/N  between slits 
are only marginal. In the same Figure we also show the skyline emission for reference.
\begin{figure*}
\epsscale{1.0}
\plotone{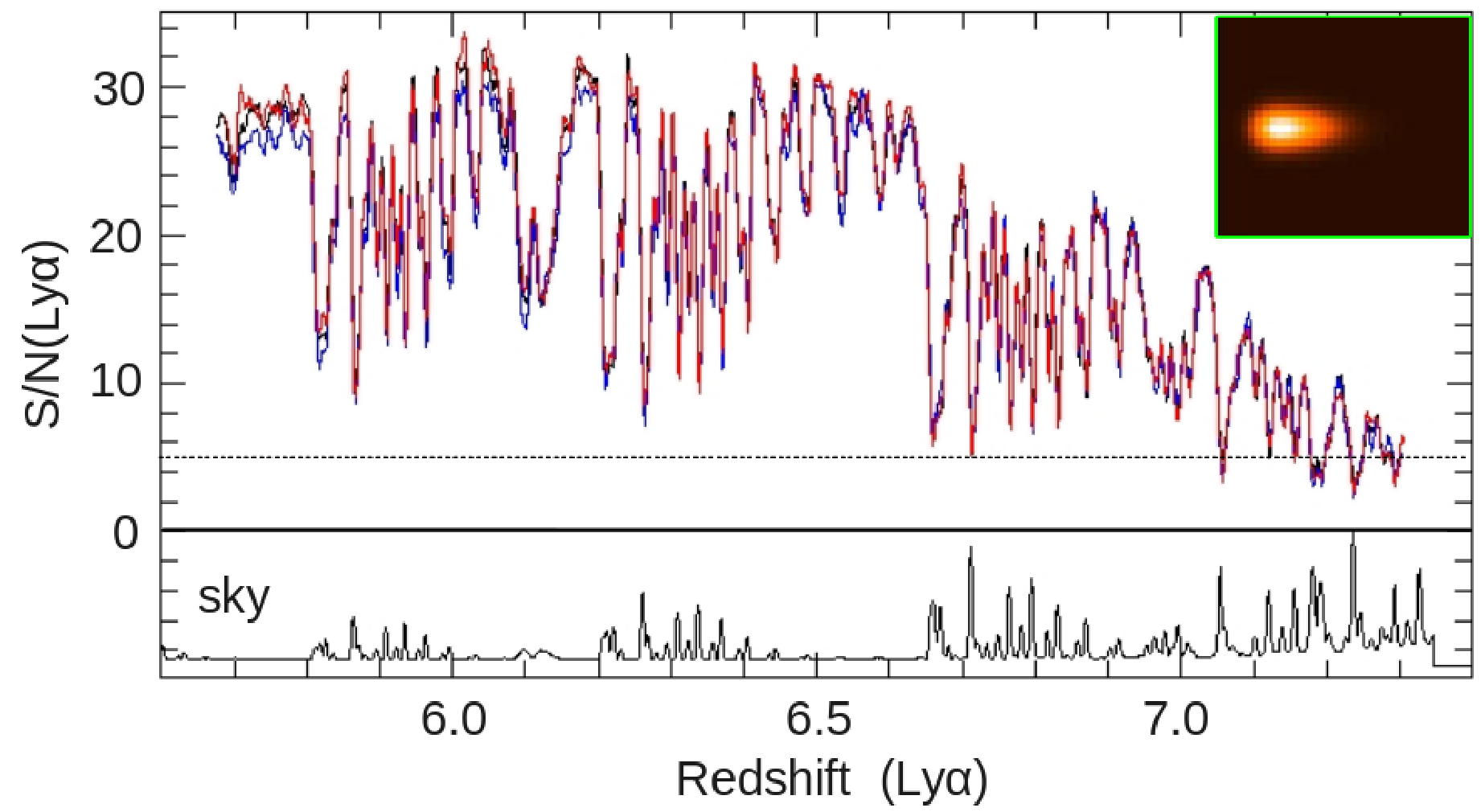}
\caption{Results from our 2D simulations. The colored curves represents the S/N 
corresponding to a line with flux $1.6e \times 10^{-17} erg/s/cm^2$ and  FWHM$=300 km/s$, at each different redshift. Each of the colors corresponds to simulations performed using a different slit in the two chips. 
For each slit the curves are the average of the S/N obtained at five different positions, respectively -10, -5, 0, +5, +10  pixels along the positional axis  with respect to the center of the slit. In the inset at the top-right corner of the figure we show a simulated Ly$\alpha$ line that is inserted in the raw frames for these simulations.
In the lower panel we show the skyline emission. }
\label{simul}
\end{figure*}

\section{The declining fraction of Ly$\alpha$ emitters: new limits and discussion}
\subsection{The fraction of Ly$\alpha$ emitters at z$\sim 7$}
With our new sample we can evaluate with greater accuracy the fraction of Ly$\alpha$ emission in LBGs at redshift 7,
 the decline between $z\sim 6$ and $z\sim 7$ and its implication.
In Table 2 we report the fraction of galaxies having and EW $> 25$ \AA\ and $> 50$ \AA\
separately for the two absolute magnitude bins that were adopted  by previous works (Stark et al. 2010, P11, Ono et al. 2012).
In the bright bin (galaxies with magnitudes  $-21.25 < M_{UV} < -20.25$) 
there are 39 galaxies of which 7 are detected in Ly$\alpha$: 5 of these have  $EW > 25$\AA\,  2 have  $EW > 50$\AA\ and none has
 $EW > 75$\AA.
In the faint bin (galaxies with  $-20.25 < M_{UV} < -18.75$) there are  25 objects,  of which 5 have a Ly$\alpha$ emission with $EW > 25$\AA\  
and 2 with  $EW > 50$\AA. Note that 3 of the targets in the Brada\v{c} et al. sample are intrinsically 
fainter than $M_{UV}=-18.75$ thus are excluded from this bin.
In the Table we report the fractions taking into account the fact that the limit in the EW 
 detectable for the  galaxies is not always below 25 \AA; 
for example for some of the objects in Ono  et al. (2012)  
the limits achieved are above this value. 
In calculating the fractions  we also consider that for  some galaxies  
the redshift probability distribution 
extends well beyond z$\sim 7.3$, which is approximately 
the limit out to which we can detect the Ly$\alpha$ emission in our 
current observations.
In particular as already stated above, the sample observed by Brada\v{c} et al. (2012) was selected in such a way that the probability 
of galaxies being at $z> 7.3$ is  quite high, $\sim 48$\% (see Figure 5 in Hall et al. 2012). This is due to the broad J-band   filter ($J110$) that was available for the selection. 
Therefore we weighted each sample by evaluating the total probability of galaxies being outside  the  redshift range that is observable by the spectroscopic setup. In practice for most of the samples 
this probability  is negligible (see Figure 6 in Ouchi et al. 2010 for the Ono et al. sample, Figure 7 in Castellano et al. 2010a for  
the NTT,GOODS-South and BDF samples), while it is non negligible for 
the UDS sample (which has a tail to z$\sim$8 , see Figure 1 in Grazian at al. 2012) and quite high for the Brada\v{c} et al. sample.
\\
We also report the  fractions after assuming that 20\% of the undetected 
objects are lower redshift interlopers: this value (20\%) is the upper limit for possible interlopers found in a large sample of z$\sim$6 galaxies in our previous work (P11), and 
we assume that there is no significant change  between the two epochs. 
Note that none of our galaxies has a detected Ly$\alpha$ emission  with EW larger than 75 \AA: to calculate the upper limit for the fraction, we assume the statistics  for small numbers of events by Gehrels (1986).

Comparing the above results to those at $z\sim 6$ presented by Stark et al. (2010), it is clear that 
there is a very significant deficit of Ly$\alpha$ emission at z$\sim 7$ compared to earlier epochs.
 We note here that very recently Schenker et al. (2014) introduced a new method to analyze the decrease of Ly$\alpha$  emission in LBGs,  based on using the 
measured slopes of the rest-frame ultraviolet continua of galaxies, rather than their absolute $M_{UV}$ 
magnitudes as we do here. According to their conclusions,  the observed difference  
between the z$\sim 6$ and z$\sim 7$ EW distribution  is even slightly larger than  with the traditional 
way of computing fractions in bins of $M_{UV}$. This is mainly because blue galaxies at $z\sim 6$ exhibit 
stronger Ly$\alpha$  emission and candidates at $z\sim 7$ tend to be bluer than at lower redshift, hence they
are expected to exhibit Ly$\alpha$ even more often.

In the following sections we will try to interpret this deficit, first within the context of 
 large-scale semi-numeric simulations of reionization that includes the reionization field as well as galactic properties
(Dijkstra et al. 2011).   We will then apply to our data a simple phenomenological model developed by Treu et
al. (2012) that uses the evolution of the distribution of Ly$\alpha$ equivalent widths to make some simple  
predictions about  the complex topology of reionization.
 
\begin{table*}
\scriptsize
\begin{center}
\caption{Fractions of Ly$\alpha$ emission}
\begin{tabular}{ccccc}
\hline
Mag  & interlopers & $EW> 25$\AA & $EW > 50$\AA & $EW > 75$\AA  \\
\hline 
\\
 $-21.25 < M_{UV} < -20.25$& none    &  0.15$^{+0.11}_{-0.08}$ & 0.06$^{+0.07}_{-0.04}$  & $< 0.05$\\
                          & 20\%    &  0.19$^{+0.13}_{-0.10}$  & 0.07$^{+0.09}_{-0.05}$ & $< 0.06$ \\
\hline
\\
$-20.25 < M_{UV} < -18.75$ & none    &  0.29$^{+0.20}_{-0.15}$  & 0.10$^{+0.13}_{-0.06}$  & $<0.08$ \\
                          & 20\%    &  0.36$^{+0.25}_{-0.18}$ & 0.12$^{+0.17}_{-0.08}$  & $<0.10$\\
\hline 
\\
all                       & none  & 0.19$^{+0.08}_{-0.06}$ & 0.07$^{+0.05}_{-0.03}$ &$<0.031$ \\
                          & 20\% & 0.23$^{+0.10}_{-0.07}$ & 0.09$^{+0.07}_{-0.04}$ & $<0.039$ \\ 

\end{tabular}
\end{center}
\footnotesize{Note: the limits  at 75\AA\ have been calculated using the confidence limits for small numbers of event (Gehrels 1986). Note that the bin with all galaxies contains also few objects which are fainter than the -18.75 limit.}
\label{table:frac} 
\end{table*}
\subsection{The neutral hydrogen fraction}
In P11 we interpreted the drop in the Ly$\alpha$ fraction in LBGs as due most probably to the
 sudden increase of neutral hydrogen  in the universe between z$\sim$6 and z$\sim 7$ (see also Schenker et al. 2012). 
We then compared  the results  to the predictions of Dijkstra et al. (2011), to determine what fraction of 
neutral hydrogen would be needed to explain the drop, provided that all other physical parameters (e.g. dust content,
the escape fraction of Lyman continuum photons etc) would not change between z=6 and z=7. We obtained a rather high neutral hydrogen fraction by volume, $\chi_{HI} \sim 0.6$.
\\
We now make use of improved models to compare our new results.
As in Dijsktra et al. (2011), reionization morphologies were generated using the public code, DexM
(Mesinger \& Furlanetto 2007). The box size is 200 Mpc, and the ionization field is
computed on a $500^3$ grid.  Reionization morphologies at a given $\chi_{HI}$ 
 are generated
by varying the ionization efficiency of halos, down to a minimum halo mass of $5\times 10^8 M_\odot$, roughly corresponding to the average minimum mass of halos at z=7 which retain enough
gas to form stars efficiently (Sobacchi \& Mesinger 2013).
Compared to P11   the  model  now includes 
also  more massive halos, with stellar masses up to $10^{12} M_\odot$.
This is because the previous model was tailored to analyze the nature of fainter dropout 
galaxies (Dijkstra et at. 2011) compared to those presented in this work.
The results however change only minimally with the new choice of
halo mass, as expected given that the halo bias (and the associated opacity
distribution for a given $\chi_{HI}$) does not evolve much over this mass range (e.g.
Mesinger \& Furlanetto 2008, McQuinn et al. 2008).
In Figure 4 we present the comparison of the outcome of the new model   
with the present fractions for the faint sample. The red circles (and limit) show the 
fractions  assuming that all our non detected target are at z$\sim7$ (the same assumption that is made in this model at z$\sim 6$) while the blue circles and limit assume  20\% interlopers.
It is clear that only a  very  high neutral hydrogen fraction ($\chi_{HI}\geq 0.51$) 
 can best reproduce  the lack of Ly$\alpha$ emission at $z=7$ compared to earlier epochs, 
even if there are still considerable uncertainties i.e. large error bars due to the 
small size of the samples. 
This high value  seems at odds with other observational results: for instance, Raskutti et al. (2012) study the IGM temperature in quasar near-zones and find that reionization must have been completed by $z > 6.5$ at high confidence,  while several teams (Hu et al. (2010), Kashikawa et al. (2011) 
and Ouchi et al. 2010)  study the Ly$\alpha$ line shapes of LAEs at z = 6.5 
 finding  no evidence of damping wings. 
\\
\begin{figure}
\epsscale{1.3}
\plotone{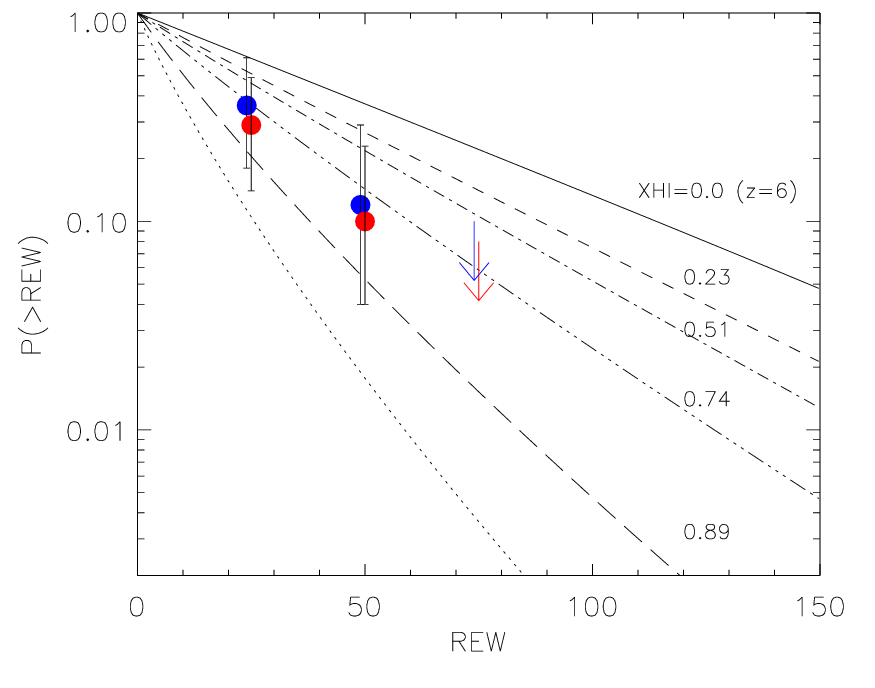}
\caption{The expected cumulative 
distribution of Ly$\alpha$ detections as a function of rest-frame EWs for z$\sim 7 $ faint  LBGs, under the 
assumption that the observed LAE fraction at $z\sim 7$ is different from z$\sim 6$ only because of the IGM. The different lines correspond to a Universe that was respectively $\sim$ 0.0, 0.23, 0.51, 0.74,0.89 and 0.94 neutral by volume
 (from top to bottom) at z=7. The line for $\chi_{HI} =0.0$ is the same as at z=6.
The lines correspond to the model with  $N_{HI}=10^{20} cm^{-2}$ and  $v_{wind}=200 km s^{-1}$.
The red(blue)   circles and limits  are our results assuming 
that 0(20)\% or the undetected galaxies are interlopers respectively (from Table 2). }
\label{marknew}
\end{figure}
As recently pointed out by  Taylor \& Lidz (2014),  before reionization completes, the simulated Ly$\alpha$
 fraction might have large spatial fluctuations depending on the degree of homogeneity/inhomogeneity of the reionization process. 
Since existing measurements of the Ly$\alpha$ fraction span relatively small regions on the sky, 
and sample these regions only sparsely (typically only a few dropouts are observed per field),
they might  by chance probe mostly galaxies with above than average Ly$\alpha$ attenuation and therefore 
point to higher neutral hydrogen fractions compared to the average values.
It is therefore important to  include the effect of cosmic variance for different sight-lines within our survey.
In their work Taylor \& Lidz found that the  sample variance is 
non negligible for existing surveys, 
and  it does somewhat mitigate the required neutral fraction at z$\sim$ 7.  
\\
Compared to previous studies and to the surveys analyzed by Taylor \& Lidz, 
this work presents 
more independent fields of view:  even considering as single pointing those in adjacent areas (e.g. the GOODS-South/ERS areas and the 
2 SDF pointings in Ono et al. 2012),  we are now sampling 8 independent lines of sight, 
with areas varying between  $50-100 {\rm arcmin}^2$ in each field.
We have verified  what is the uncertainty in our results that might be derived from the
cosmic variance.

 We have tried to quantify what is the variance in the opacity ($e^{-\tau}_{reion}$) in the simulations, 
due  to  the limited number of fields analyzed.  
In the  simulations  we take 8 random regions with areas corresponding to our observed fields 
and  in each field we sample a reasonable number of lines-of-sight (corresponding to the average 
number of candidates spectroscopically observed).  We then  compute the pointing-to-pointing (cosmic) standard deviation $\sigma_{FOV}(e^{-\tau}) $, which for 8 field is of the order of 
6\%. The variance is so small because the total volume sampled is quite large.
We also varied the number of candidate galaxies probing the reionization 
per each pointing: indeed the 
discreteness  in sampling the reionization morphology becomes an
issue  especially for large  neutral fractions 
 since in this case  the fewer number of would-be Ly$\alpha$ emitters  can miss the relatively
rare regions in a given pointing that have a high transmission.
However even considering a small sampling (only 5 emitters per pointing) $\sigma_{FOV}(e^{-\tau}) $
is still around 10\%.  
 Note that this  computation just uses the opacity at a
single wavelength, roughly 200 $km s^{-1}$ red-ward of the line center, where most of the
intrinsic emission is expected to lie. This is not really the variance in the Ly$\alpha$ fraction, 
since the latter requires some more detailed modeling, but it gives a crude idea of what is the expected cosmic variance from reionization.
Therefore we are quite confident that for our sample  which has  a large number of independent field 
and a reasonable number of candidates observed per pointing, the field to field fluctuations are not very large and would not effect sensibly the results of Figure 4.

\subsection{A patchy reionization process?}
Applying the simple phenomenological models developed by Treu et
al. (2012) to describe the evolution of the distribution of Ly$\alpha$
equivalent widths
we can now use the Ly$\alpha$ detections and non-detections
to make some inferences about the complex topology of reionization.
This model starts from the intrinsic rest-frame distribution in terms
of the one measured at $z \sim 6$ by Stark et al. (2011).  It then
considers two extreme cases which should bracket the range of possible
scenarios for the reionization morphology: in the first ("patchy")
model, no Ly$\alpha$ is received from a fraction $\epsilon_p$ of the
sources, while the rest is unaffected. In the second ("smooth") model,
the Ly$\alpha$ emission is attenuated in every galaxy in the same way,
by a factor $\epsilon_s$.  These two models can be thought
respectively as simple idealizations of smooth and patchy
reionization: although very simple and somewhat unphysical (especially
the smooth one) these two models should bracket the expected
behavior of the IGM near the epoch of reionization (see Treu et
al. 2012, Treu et al. 2013 for a more detailed explanation).
\\
For each object in our sample,  Bayes's rule gives the posterior probability of $\epsilon_p$ and $\epsilon_s$ (which we collectively indicate as $\epsilon$)  and redshift  given  the  observed spectrum and the continuum magnitude.  The likelihood is as usual the probability of obtaining the data for any given value of the parameter.
The model adopts a uniform prior p($\epsilon$) between zero and unity, while the prior for the redshift $p(z)$  is obtained from  the  redshift probability distribution (as  described in Section 3 for each of the different parent samples). 
We use the implementation of the method that takes as input the line equivalent widths or equivalent width limits, in order to incorporate all the available information, even when a noise spectrum is not available (Treu et al. 2012).

One of the output of the model is the normalization constant $Z$,
known as the Bayesian evidence and quantifies how well each of the two
models matches the data.  The evidence ratio is a powerful way to
perform model selection e.g., comparing the patchy and smooth models
and eventually discriminate between the two.  
%
\begin{figure}
\epsscale{1.2}
\includegraphics[width = 9cm,clip=]{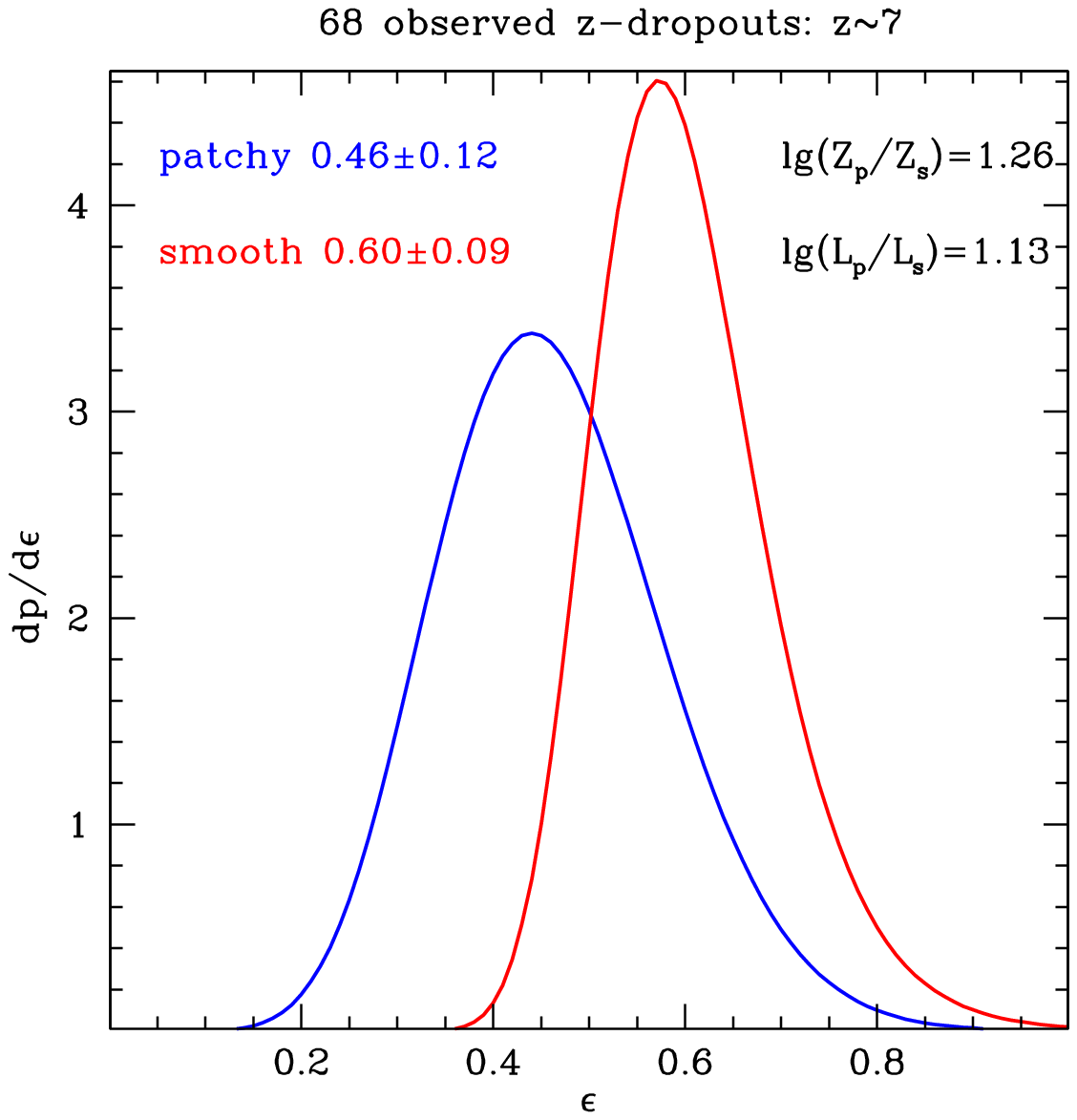}
\caption{
 Marginalized posterior distribution function of Ï$\epsilon$ at z $\sim$ 7 based on a compilation of 68 $z$-dropouts with deep spectroscopic follow-up presented in this paper or taken from the literature (Pentericci et al. 2011; Ono et al. 2012; Schenker et al. 2012). Both the patchy and smooth model indicate clearly that the Ly$\alpha$ emission is significantly quenched at z $\sim$ 7 with respect to z $\sim$ 6 (i.e. $\epsilon<1$).}
\label{prob}
\end{figure}

\begin{figure}
\epsscale{1.2}
\includegraphics[width = 9cm]{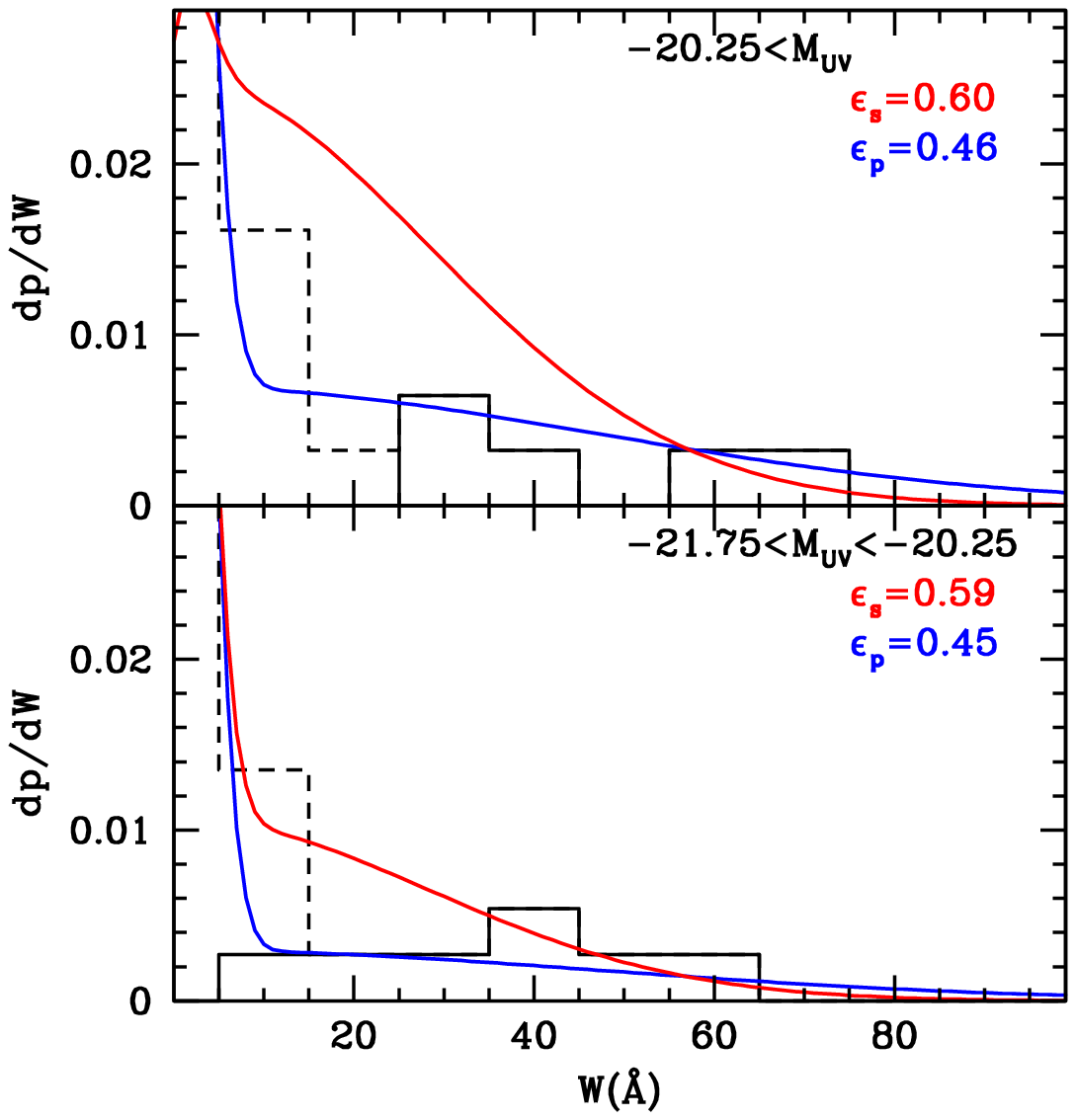}
\caption{
  The predicted distribution of rest-frame equivalent width  for the best 
patchy (blue) and smooth (red) models, for the bright sub-sample (top panel) and 
faint  sub-samples (lower panel). The black histograms are based on the detected
Ly$\alpha$ emitters in each sample, 
while   the dashed ones  assume that all undetected objects have a Ly$\alpha$ 
that is immediately below  the threshold (so they represent upper limits).
}
\label{histo}
\end{figure}

Treu et al. 2012 applied their model to the sample presented by Ono et
al. (2012) which included also data from P11 and V11.  The data
clearly preferred an attenuation factor $\epsilon < 1$ (0.65-0.68),
independent of the model considered, but the evidence ratio indicated
no significant preference for any of the two models.
\\
We have repeated the exercise for our new enlarged sample, which is
almost double compared to the previous one and most importantly
contains a larger fraction of very faint galaxies (thanks e.g. to the
inclusion of the lensed galaxies of the Bradac et al. sample, and several other faint targets 
 from UDS and the archival data) .  
With the new sample we obtain
$\epsilon_p =0.46 \pm 0.12$ and $\epsilon_s =0.60 \pm 0.09$
respectively for the patchy and smooth model, as shown in Figure
\ref{prob}: this means that both models require a considerable
quenching of the Ly$\alpha$ compared to z$\sim 6$, as expected from
the many non-detections.  Note that these results assume that the
level of contamination in the samples, is the same at $z=6$ and at $z=7$.
We can interpret the $\epsilon_p$ and $\epsilon_s$ as the average
excess optical depth of Ly$\alpha$ with respect to z $\sim$6, i.e.,
$\langle e^{-\tau _{\rm Ly\alpha }}\rangle$, although a conversion
from this to a neutral hydrogen fraction requires detailed and
uncertain modeling (e.g. Santos 2004).  A key result is that with the
new sample the evidence ratio between the two models is quite high,
$log(Z_p/Z_s )=1.26$ which means that the patchy model is highly
favored ($>18$ times) by the data over the smooth one.  This is
also suggested by the likelihood ratio test: $L_p/L_s$ strongly favors
the patchy model (in other terms the ratio corresponds to a difference
in $\Delta \chi^2$ between the two models of $2 ln (L_p/L_s) \sim
5.2$). As expected the power to discriminate between the two models is
given by the inclusion of fainter galaxies, as well as the fact that
for many galaxies we have very deep EW limits.  As a result of the
inference, the model also allows us to calculate the fraction of
emitters using all the available information.
For objects brighter than $M_{UV}=-20.25$ the model predicts respectively 
$0.09\pm0.04$ for galaxies with $EW>25$ \AA\ and  $0.03\pm0.02$  for galaxies with $EW>55$ \AA.  For fainter galaxies, the predictions are $0.24\pm0.08$ and       $0.12\pm0.05$ respectively. These values are very close to the numbers reported in Table 2 (considering obviously the fractions derived assuming no interlopers in the sample).
In Figure \ref{histo} we show the predicted distribution of rest-frame equivalent width  for the best 
patchy (blue) and smooth (red) models, for the bright and faint sub-sample separately. 
The black histograms are based on the detected
Ly$\alpha$ emitters in each sample. In particular the blue model makes predictions that are 
 closer to the real data for the faint sub-sample, since it predicts better the high EW tail and it does not show a deficit of detections at intermediate EW ($\sim 30-40$ \AA) 
While our observational  results indicate clearly that the distribution of neutral hydrogen in these phases of reionization was 
highly inhomogeneous, as expected by most theoretical predictions 
(e.g. Iliev et al. 2006),  to fully constrain the morphology of reionization we will have to wait for the
direct observations of the 21 cm emission from neutral hydrogen in the
high redshift universe, which is one of the prime tasks of the
upcoming LOFAR surveys observations (e.g. Jensen et al. 2013).

\section{Summary and concluding remarks}

In this paper we have presented new results from our search for z$\sim
7 $ galaxies from deep spectroscopic observations of candidate
z-dropouts in the CANDELS fields.  Even though our sensitive VLT
observations reached extremely low flux limits, only two galaxies
have new robust redshift identifications, one from the Ly$\alpha$
emission line at z=6.65 and the other from its Lyman-alpha break, i.e. the
continuum discontinuity at the Ly$\alpha$ wavelength consistent with a
redshift $\sim$6.42. In this second object no emission line is
observed. In addition for $23$ galaxies we present new deep limits on
the Ly$\alpha$ EW derived from the non detections in ultra-deep
observations (from 15 to 27 hours) obtained with FORS2 spectrograph on
the VLT.  Using this new data as well as previously published samples,
we have assembled a total of 68 candidate z$\sim 7 $ galaxies with
deep spectroscopic observations, of which 12 have a redshift
identification from the Ly$\alpha$ emission line. With this much enlarged sample we have placed solid
constraints on the fraction of Ly$\alpha$ emission in z$\sim$7 Lyman
break galaxies both for bright and faint galaxies, confirming the
large decline in the presence of Ly$\alpha$ emission from z$\sim 6$ to
$z\sim 7$. If this decline is only due to the evolution of the IGM,
and assuming that all other galaxies properties remain unchanged in
this redshift interval, a very large fraction ($\chi_{HI}\geq
0.51$) of neutral hydrogen is needed to explain the observations.  Finally
applying the simple phenomenological model developed by Treu et
al. (2012), we show that the present data favor a patchy reionization
process rather than a smooth one, as expected from most simulations
 (e.g. Friedrich et al. 2011, Choudhury et al. 2009, Iliev et al. 2006 to name a few: also see Trac \& Gnedin 2009 for a review on simulations of reionization).

Obviously we cannot rule out that an evolution of other properties, namely $f_{esc}$ and
dust, come into play and contribute to the Ly$\alpha$ quenching\footnote{The
 Lya EW-PDF is affected by a combination of $f_{esc}$, dust and IGM
transmission. Existing observations at a fixed redshift cannot break
degeneracies between these different parameters (as shown clearly by
 Hutter et al. 2014). However, there is additional information in the observed
 redshift evolution which can place more stringent constraints (see e.g.
Dijkstra et al. 2014).} Indeed, in  a recent paper  (Dijkstra et al. 2014) we discuss the possibility  that the decline in strong Ly$\alpha$ emission from $z > 6$ galaxies is due, in part, also to an increase of the Lyman continuum escape fraction in star forming galaxies. In particular assuming that the escape fraction evolves with redshift as 
$f_{\it esc}(z) = f_0 ([1 + z]/5)^k$ (as in Kuhlen \& Faucher-Giguere
2012), and taking $k=4$ and $f_0 =0.04$ such that  we have $f_{\it esc} = 0.15$ at z=6 and $f_{\it esc} = 0.26$ at z=7,  the observed decline in Ly$\alpha$ emission could be reproduced  with a more modest  evolution in the global neutral fraction, of the order of $\Delta \chi_{HI} \sim 0.2$. 
This work is clearly rather speculative,  since $f_{\it esc}$ is a very elusive quantity 
to measure and we only have tentative indications on its  value 
from upper limits  (e.g. Nestor et al. 2013, Vanzella et al. 2012, Boutsia et al. 2011) and on its evolution (e.g. Cowie et al. 2009, Siana et al. 2010). However  it shows that  an evolving escape fraction of  ionizing photons should be considered  as part of the  explanation for evolution in the
Ly$\alpha$  emission of high redshift galaxies in addition to the evolution of the IGM 
(see Dijkstra et al. 2014 for more details). 

It is clear that to fully characterize and understand the reionization epoch, and to clarify the relation 
between the disappearing  Ly$\alpha$  emission line and cosmic reionization
 we still  have to make a substantial  effort,  both on the observational and on the modeling side. 
Even if the current  samples of candidate
galaxies at $z > 7$ are quite large, and despite all the observational efforts by several teams, 
the number of spectroscopically confirmed objects remains very small. 
To overcome this limitation and  substantially  increase the statistics  we are currently  carrying out an ESO Large Program with FORS2@VLT (PI Pentericci) that in the end   should allow us to increase  considerably  the number of  confirmed  high redshift galaxies by observing $\sim 200$  candidates.
 In particular, since the targets will be selected from the CANDELS field with extremely deep near-IR observations, we will include  also  galaxies as faint as  $J=27$. 
 The new deep spectroscopic observations  will allow us to 
assess the  continuous evolution  of the Ly$\alpha$ emission  over the range $6 < z < 7.3$,  and  from a comparison  to state-of-the art models, we will be able to  determine if  the Ly$\alpha$ was  mainly quenched by the neutral IGM, or if any evolution of the galaxies' physical properties also played a significant role.

\acknowledgments
Part of this work has been funded through INAF Grants (PRIN-INAF  2010 and 2012).RJM acknowledges the support of the European Research Council via the award of a Consolidator Grant 
JSD acknowledges the support of the European Research
Council via the award of an Advanced Grant,
the support of the Royal Society via a Wolfson Research Merit
Award, and the contribution of the EC FP7 SPACE
project ASTRODEEP (Ref.No: 312725).


\begin{thebibliography}{}
\expandafter\ifx\csname natexlab\endcsname\relax\def\natexlab#1{#1}\fi


\bibitem[Bolton 
\& Haehnelt(2013)]{2013MNRAS.429.1695B} Bolton, J.~S., \& Haehnelt, M.~G.\ 2013,  \mnras, 429, 1695 

\bibitem[Boutsia et al.(2011)]{2011ApJ...736...41B} Boutsia, K., Grazian, 
A., Giallongo, E., et al.\ 2011, \apj, 736, 41 

\bibitem[Bouwens et al.(2011)]{2011ApJ...737...90B} Bouwens, R.~J., 
Illingworth, G.~D., Oesch, P.~A., et al.\ 2011, \apj, 737, 90 

\bibitem[Bowler et al.(2014)]{2014MNRAS.440.2810B} Bowler, R.~A.~A., 
Dunlop, J.~S., McLure, R.~J., et al.\ 2014, \mnras, 440, 2810 


\bibitem[Brada{\v c} et al.(2012)]{2012ApJ...755L...7B} Brada{\v c}, M., 
Vanzella, E., Hall, N., et al.\ 2012, \apjl, 755, L7 

\bibitem[Caruana et al.(2012)]{car12} Caruana, J., Bunker, 
A.~J., Wilkins, S.~M., et al.\ 2012, \mnras, 427, 3055 

\bibitem[Caruana et al.(2013)]{caru2013} Caruana, J., Bunker, 
A.~J., Wilkins, S.~M., et al.\ 2013, arXiv:1311.0057 

\bibitem[Castellano et 
al.(2010)]{2010A&A...524A..28C} Castellano, M., Fontana, A., Paris, D., et al.\ 2010b, \aap, 524, A28 

\bibitem[Castellano et 
al.(2010)]{2010A&A...511A..20C} Castellano, M., Fontana, A., Boutsia, K., et al.\ 2010a, \aap, 511, A20 

\bibitem[Choudhury et al.(2009)]{2009MNRAS.394..960C} Choudhury, T.~R., 
Haehnelt, M.~G., \& Regan, J.\ 2009, \mnras, 394, 960 


\bibitem[Cl{\'e}ment et 
al.(2012)]{2012A&A...538A..66C} Cl{\'e}ment, B., Cuby, J.-G., Courbin, F., et al.\ 2012, \aap, 538, A66 

\bibitem[Cowie et al.(2009)]{2009ApJ...692.1476C} Cowie, L.~L., Barger, 
A.~J., \& Trouille, L.\ 2009, \apj, 692, 1476 

\bibitem[Dijkstra et al.(2014)]{2014arXiv1401.7676D} Dijkstra, M., Wyithe, 
S., Haiman, Z., Mesinger, A., \& Pentericci, L.\ 2014, arXiv:1401.7676 

\bibitem[Dijkstra et al.(2011)]{2011MNRAS.414.2139D} Dijkstra, M., 
Mesinger, A., \& Wyithe, J.~S.~B.\ 2011, \mnras, 414, 2139 

\bibitem[Faisst et al.(2014)]{faisst} Faisst, A.~L., Capak, 
P., Carollo, C.~M., Scarlata, C., \& Scoville, N.\ 2014, arXiv:1402.3604 

\bibitem[Finkelstein et al.(2012)]{2012ApJ...756..164F} Finkelstein, S.~L., 
Papovich, C., Salmon, B., et al.\ 2012, \apj, 756, 164 

\bibitem[Fontana et al.(2010)]{2010ApJ...725L.205F} Fontana, A., Vanzella, 
E., Pentericci, L., et al.\ 2010, \apjl, 725, L205 

\bibitem[Forero-Romero et al.(2012)]{2012MNRAS.419..952F} Forero-Romero, 
J.~E., Yepes, G., Gottl{\"o}ber, S., \& Prada, F.\ 2012, \mnras, 419, 952 

\bibitem[Friedrich et al.(2011)]{2011MNRAS.413.1353F} Friedrich, M.~M., 
Mellema, G., Alvarez, M.~A., Shapiro, P.~R., 
\& Iliev, I.~T.\ 2011, \mnras, 413, 1353 


\bibitem[Furlanetto 
\& Oh(2005)]{2005MNRAS.363.1031F} Furlanetto, S.~R., \& Oh, S.~P.\ 2005, \mnras, 363, 1031 

\bibitem[Furlanetto et al.(2006)]{2006MNRAS.365..115F} Furlanetto, S.~R., 
McQuinn, M., \& Hernquist, L.\ 2006, \mnras, 365, 115 

\bibitem[Galametz et al.(2013)]{2013ApJS..206...10G} Galametz, A., Grazian, 
A., Fontana, A., et al.\ 2013, \apjs, 206, 10 

\bibitem[Gehrels(1986)]{1986ApJ...303..336G} Gehrels, N.\ 1986, \apj, 303, 
336 

\bibitem[Grazian et 
al.(2012)]{2012A&A...547A..51G} Grazian, A., Castellano, M., Fontana, A., et al.\ 2012, \aap, 547, A51 

\bibitem[Hall et al.(2012)]{2012ApJ...745..155H} Hall, N., Brada{\v c}, M., 
Gonzalez, A.~H., et al.\ 2012, \apj, 745, 155 

\bibitem[Hu et al.(2010)]{2010ApJ...725..394H} Hu, E.~M., Cowie, L.~L., 
Barger, A.~J., et al.\ 2010, \apj, 725, 394 

\bibitem[Iliev et al.(2006)]{2006MNRAS.369.1625I} Iliev, I.~T., Mellema, 
G., Pen, U.-L., et al.\ 2006, \mnras, 369, 1625 

\bibitem[Jensen et al.(2013)]{2013MNRAS.428.1366J} Jensen, H., Laursen, P., 
Mellema, G., et al.\ 2013, \mnras, 428, 1366 

\bibitem[Jensen et al.(2013)]{2013MNRAS.435..460J} Jensen, H., Datta, 
K.~K., Mellema, G., et al.\ 2013, \mnras, 435, 460 

\bibitem[Kashikawa et al.(2011)]{2011ApJ...734..119K} Kashikawa, N., 
Shimasaku, K., Matsuda, Y., et al.\ 2011, \apj, 734, 119 

\bibitem[Loeb 
\& Rybicki(1999)]{loeb} Loeb, A., \& Rybicki, G.~B.\ 1999, \apj, 524, 527 

\bibitem[Malhotra 
\& Rhoads(2006)]{malho} Malhotra, S., \& Rhoads, J.~E.\ 2006, \apjl, 647, L95 

\bibitem[McLure et al.(2010)]{2010MNRAS.403..960M} McLure, R.~J., Dunlop, 
J.~S., Cirasuolo, M., et al.\ 2010, \mnras, 403, 960 

\bibitem[McQuinn et al.(2007)]{2007MNRAS.377.1043M} McQuinn, M., Lidz, A., 
Zahn, O., et al.\ 2007, \mnras, 377, 1043 

\bibitem[Mesinger 
\& Furlanetto(2008)]{2008MNRAS.386.1990M} Mesinger, A., \& Furlanetto, S.~R.\ 2008, \mnras, 386, 1990 

\bibitem[Mesinger 
\& Furlanetto(2007)]{2007ApJ...669..663M} Mesinger, A., \& Furlanetto, S.\ 2007, \apj, 669, 663 

\bibitem[Nestor et al.(2013)]{2013ApJ...765...47N} Nestor, D.~B., Shapley, 
A.~E., Kornei, K.~A., Steidel, C.~C., \& Siana, B.\ 2013, \apj, 765, 47 

\bibitem[Ono et al.(2012)]{2012ApJ...744...83O} Ono, Y., Ouchi, M., 
Mobasher, B., et al.\ 2012, \apj, 744, 83 

\bibitem[Ota et al.(2008)]{2008ApJ...677...12O} Ota, K., Iye, M., 
Kashikawa, N., et al.\ 2008, \apj, 677, 12 


\bibitem[Ouchi et al.(2010)]{2010ApJ...723..869O} Ouchi, M., Shimasaku, K., 
Furusawa, H., et al.\ 2010, \apj, 723, 869 

\bibitem[Pentericci et al.(2011)]{2011ApJ...743..132P} Pentericci, L., 
Fontana, A., Vanzella, E., et al.\ 2011, \apj, 743, 132 

\bibitem[Raskutti et al.(2012)]{2012MNRAS.421.1969R} Raskutti, S., Bolton, 
J.~S., Wyithe, J.~S.~B., \& Becker, G.~D.\ 2012, \mnras, 421, 1969 

\bibitem[Santos(2004)]{2004MNRAS.349.1137S} Santos, M.~R.\ 2004, \mnras, 
349, 1137 

\bibitem[Schenker et al.(2012)]{she12} Schenker, M.~A., 
Stark, D.~P., Ellis, R.~S., et al.\ 2012, \apj, 744, 179 

\bibitem[Schenker et al.(2014)]{2014arXiv1404.4632S} Schenker, M.~A., 
Ellis, R.~S., Konidaris, N.~P., \& Stark, D.~P.\ 2014, arXiv:1404.4632 

\bibitem[Schmidt et al.(2014)]{2014arXiv1402.4129S} Schmidt, K.~B., Treu, 
T., Trenti, M., et al.\ 2014, arXiv:1402.4129 

\bibitem[Siana et al.(2010)]{2010ApJ...723..241S} Siana, B., Teplitz, 
H.~I., Ferguson, H.~C., et al.\ 2010, \apj, 723, 241 

\bibitem[Sobacchi 
\& Mesinger(2014)]{2014arXiv1402.2298S} Sobacchi, E., \& Mesinger, A.\ 2014, arXiv:1402.2298 

\bibitem[Sobacchi 
\& Mesinger(2013)]{2013MNRAS.432.3340S} Sobacchi, E., \& Mesinger, A.\ 2013, \mnras, 432, 3340 

\bibitem[Stark et al.(2011)]{2011ApJ...728L...2S} Stark, D.~P., Ellis, 
R.~S., \& Ouchi, M.\ 2011, \apjl, 728, L2 

\bibitem[Stark et al.(2010)]{2010MNRAS.408.1628S} Stark, D.~P., Ellis, 
R.~S., Chiu, K., Ouchi, M., \& Bunker, A.\ 2010, \mnras, 408, 1628 

\bibitem[Taylor 
\& Lidz(2013)]{2013arXiv1308.6322T} Taylor, J., \& Lidz, A.\ 2013, arXiv:1308.6322 

\bibitem[Trac 
\& Gnedin(2011)]{2011ASL.....4..228T} Trac, H.~Y., \& Gnedin, N.~Y.\ 2011, Advanced Science Letters, 4, 228 

\bibitem[Treu et al.(2013)]{2013ApJ...775L..29T} Treu, T., Schmidt, K.~B., 
Trenti, M., Bradley, L.~D., \& Stiavelli, M.\ 2013, \apjl, 775, L29 

\bibitem[Treu et al.(2012)]{2012ApJ...747...27T} Treu, T., Trenti, M., 
Stiavelli, M., Auger, M.~W., \& Bradley, L.~D.\ 2012, \apj, 747, 27 

\bibitem[Vanzella et al.(2012)]{2012ApJ...751...70V} Vanzella, E., Guo, Y., 
Giavalisco, M., et al.\ 2012, \apj, 751, 70 

\bibitem[Vanzella et al.(2011)]{2011ApJ...730L..35V} Vanzella, E., 
Pentericci, L., Fontana, A., et al.\ 2011, \apjl, 730, L35 

\bibitem[Wilkins et al.(2011)]{2011MNRAS.411...23W} Wilkins, S.~M., Bunker, 
A.~J., Lorenzoni, S., \& Caruana, J.\ 2011, \mnras, 411, 23 

\bibitem[Wilkins et al.(2014)]{2014MNRAS.439.1038W} Wilkins, S.~M., 
Stanway, E.~R., \& Bremer, M.~N.\ 2014, \mnras, 439, 1038 


\bibitem[Zheng et al.(2010)]{zheng} Zheng, Z., Cen, R., Trac, 
H., \& Miralda-Escud{\'e}, J.\ 2010, \apj, 716, 574 



\end{thebibliography}
\end{document}